# A Study on Improvement of Image Quality in Quantum Polarized Microscopy using an Entangled-Photon Source

Mousume Samad [1,2]*, Maki Shimizu [1] and Yasuto Hijikata [1]*

[1] *Department of Electrical and Electronic Systems Engineering, Faculty of Engineering, Saitama University, 255 Shimo-Okubo, Saitama-shi, Saitama 338-8570, Japan*

[2] *Department of Information and Communication Engineering, Bangladesh Army University of Engineering and Technology (BAUET), Qadirabad 6431, Bangladesh*

**\*** Correspondence: samad.m.128@ms.saitama-u.ac.jp; yasuto@mail.saitama-u.ac.jp

**Abstract**

Quantum imaging is an advanced method for microscopy or investigating the optical properties of materials or bio-medical inspections with high accuracy, low noise, and extremely low photo-damage. In previous work, we proposed a quantum imaging technique, in which variations in polarization angle due to reflection or transmission at the sample enhance image contrast in the second-order autocorrelation function at zero delay $g^2(0)$. However, the image resolution observed was not satisfactory. In this report, to improve the image quality, the parameters such as coincidence window was optimized to distinguish photon statistics effectively due to change in polarization angle and the objective lens was utilized to focus the entangled photon beams. These modifications enabled optical imaging by accurately detecting slight polarization variations in different media. It is suggested that the proposed method enables a superior imaging performance even in the presence of noise and low-power illumination requirement, offering new possibilities for high-quality imaging applications.

## 1. Introduction

Optical microscopy has been fundamental tools for driving advancements in research in material science, biology, and other fields that depend on high-resolution imaging [1]. In this field, the progress in understanding light properties has consistently inspired the development of groundbreaking imaging technologies. In particular, non-classical light sources, such as single-photon and entangled-photon sources, have enabled quantum microscopy techniques that overcome the limitations of traditional optical imaging with classical light in terms of signal-to-noise ratio, resolution, contrast, and spectral range [1-2]. Additionally, quantum imaging offers exceptional robustness against external disturbances, including mechanical vibrations, environmental fluctuations, and variations in the light source [3-7].

In biological observations, we often face challenges, particularly the trade-off between resolution enhancement and minimizing photo-damage to delicate biological structures. Namely, while increasing optical power improves resolution, excessive power can cause photo bleaching. Conversely, reducing illumination power minimizes noise susceptibility but degrades image quality. This inherent trade-off between resolution and optical power presents a significant limitation. In contrast, the entangled photons[8-9], with their unique quantum properties, offer unprecedented opportunities for achieving super-resolution and improving contrast in imaging. Entangled-photon sources enable imaging techniques in spectral ranges where efficient detection is difficult or even allow imaging with light that does not directly interact with the sample [10-13]. Moreover, quantum polarization microscopy utilizing an entangled-photon source generated through spontaneous parametric down-conversion (SPDC) [14] offers a specific quantum state of light, where their photon statistics enable sensing and imaging including operation at extremely low light intensities. This approach might open new possibilities for photosensitive biomedical applications by enabling imaging at minimal light levels, offering valuable insights into photoactive biological processes.

Ono *et al.* [15] introduces an optical microscopy technique that enables optical phase measurement beyond the standard quantum limit (SQL). This method achieved a high signal-to-noise ratio (SNR), where two photon interface enhances the measurement precision[16,17]. However, a significant drawback of this approach is the need for optical path length correction within tens of micrometers[15]. Likewise, quantum holography utilizing polarization entanglement[18] faces a major challenge, as image acquisition requires ten hours due to the limited frame rate of EMCCD cameras. To overcome the above issues, we proposed a quantum polarization microscopy technique incorporating an entangled photon source[14]. This method enables image construction with extremely low light intensity in the picojoule range[14], making it highly suitable for imaging photosensitive biochemical applications while maintaining robustness and practicality without requiring precise optical alignment. Unlike classical polarization microscopy, this technique utilizes coincidence counting of entangled photons to enhance image contrast responsible for variations in polarization angle caused by reflection or transmission. It also leverages the second-order autocorrelation function of thermal photon statistics to achieve superior sensitivity and contrast in imaging. Basically, the photon statistics of the entangled photon source is in accordance with the thermal state, similar to the blackbody radiation. However, the peak intensity in the second-order autocorrelation function at zero delay exhibits significantly higher than that obtained from the ordinary thermal state, i.e. super-Poisson state, because the photon packet usually contains two photons though their interval is relatively long. Accordingly, we called this photon state as the hyper-Poisson state, to distinguish from the super-Poisson state. When the coincidence counts are taken from either the signal or the idler photon only, the photon state remains thermal but becomes more dispersive, resulting in a lower peak intensity of the autocorrelation function, i.e., in the super-Poisson state. These distinct thermal states hyper-Poisson and super-Poisson can be switched by slightly changing the photon polarization, which undergoes a sudden transition within a narrow range of the analyzer angle [14]. Based on this methodology, we initially demonstrated the feasibility of quantum polarization microscopy[14]. However, the image quality was not satisfactory in terms of resolution and clarity.

In this report, to improve image quality, we investigated how the coincidence window affects the transition between hyper-Poisson and super-Poisson states, which in turn enables accurate media determination in the presence of noise. Additionally, objective lenses were placed at front- and back-side of the sample to improve the resolution of image. Moreover, we validated that the photon statistics is switched by rotating the analyzer. Ultimately, this study aims to establish quantum polarization microscopy as a superior imaging tool, opening new frontiers for high-quality, and low-illumination power imaging applications.

## 2. Experimental methods

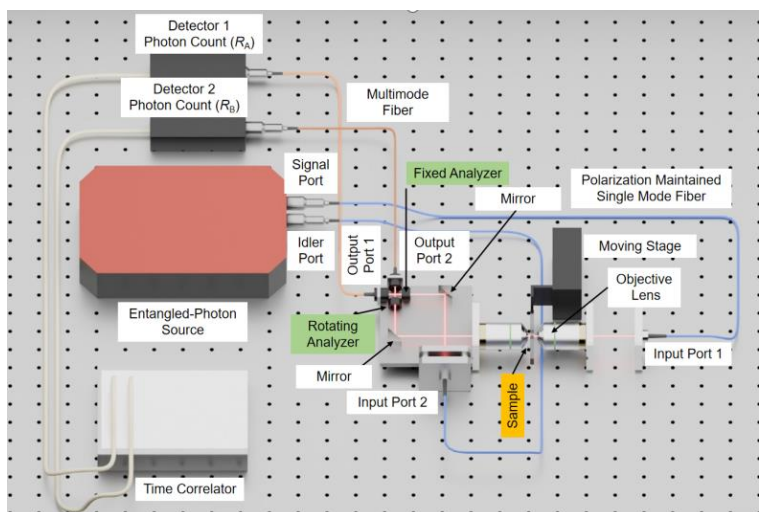

**Fig. 1.** Schematic view of the imaging system.

The schematic layout of the quantum polarization microscopy setup used in this study is illustrated in Fig.1. This experimental setup is largely identical to that in Samad *et al.*[14]. However, in this experiment, a pair of objective lenses is placed around the sample to improve the resolution of image. Whereas in these advantages, this approach may reduce photon throughput due to the lower spatial coherence of entangled photons compared to laser light.

To demonstrate the advantages of the proposed imaging method, we utilize a first-order ($m = 1$) optical spiral retarder sample[14], a USAF birefringence resolution target (Thorlabs R3L1S1B) sample, and a scotch tape sample consisting of three regions: non-tensioned scotch tape, tensioned scotch tape, and a bare glass substrate. Next section, we will show the experimental results.

## 3. Results and discussion

### 3.1 Effect of coincidence window

In previous work[14], we gave a coincidence window ($\Delta t$; 15 ns) for coincidence and autocorrelation imaging, corresponding to the width of bunching peak seen in the second-order autocorrelation vs. time delay ($g^2(\tau)$) measurement[14]. However, in order to improve image quality and reduce noise, it is necessary to optimize the coincidence window. Accordingly, we conducted $g^2(\tau)$ measurements at different coincidence windows ($\Delta t$).

The effects of different coincidence window for $g^2(\tau)$ value are illustrated in Fig. 2. With longer coincidence windows, fewer points and a smaller peak appear within the bunching peak. Additionally, the maximum of the autocorrelation function does not occur at zero delay but slightly shifts from zero. Conversely, noise is generally reduced. A larger coincidence window[19-24] results in wider time intervals for counting photon coincidences, causing multiple photon events to be grouped into single bins. This broadens the peak and reduces the number of discrete points representing it, making the time interval too coarse to capture the fine structure of the correlation peak[25]. Conversely, decreasing the coincidence window leads to the sharper bunching peak in the autocorrelation function. However, since the reduction of the coincidence window reduces the number of photon events counted per bin, leading to greater statistical fluctuations. This is particularly evident when the photon flux is low, as the signal-to-noise ratio decreases with decreasing the bin widths[26]. The standard deviation, maximum count rate, and the full width at half maximum (FWHM) obtained from curve fitting for the hyper-Poisson and super Poisson states across different coincidence windows are presented in Table 1.

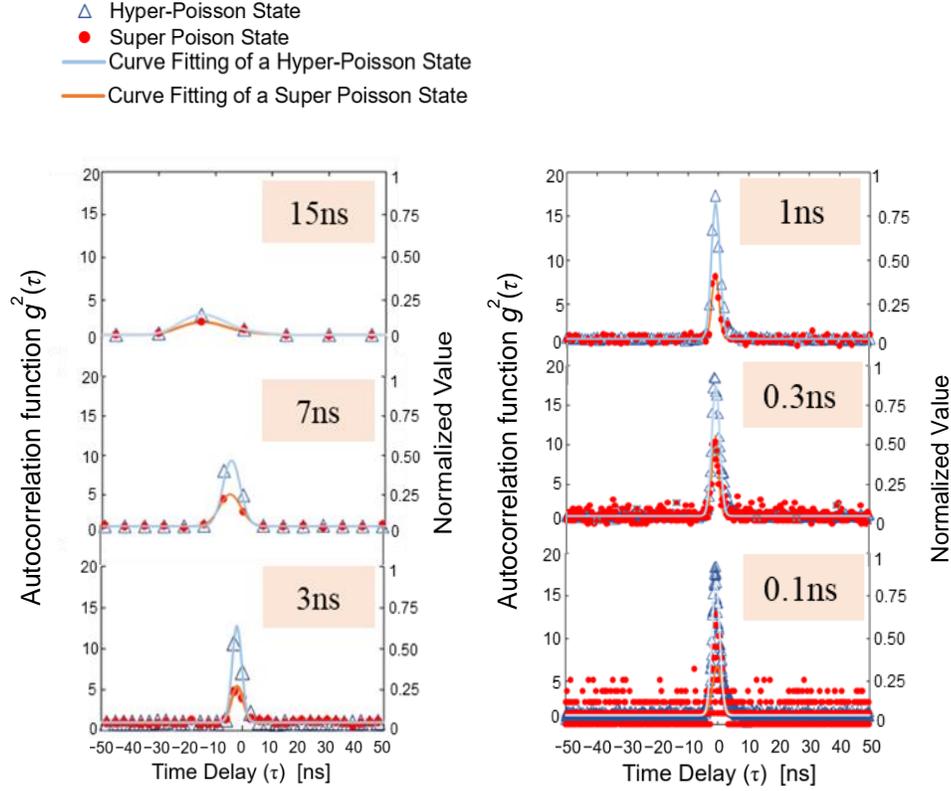

**Fig. 2.** Experimental results of $g^2(\tau)$ versus time delay ($\tau$) for hyper-Poisson state and super Poisson state at different coincidence windows.

Table 1. Standard deviation maximum count rate, and the full width at half maximum (FWHM) from curve fits for hyper-Poisson and super-Poisson states across different coincidence windows.

| Coincidence Window | Standard Deviation (%) | | Maximum Count rate [Count per bins] | | FWHM [°] | |
|---|---|---|---|---|---|---|
| | Hyper-Poisson State | Super Poisson State | Hyper-Poisson State | Super Poisson State | Hyper-Poisson State | Super Poisson State |
| 15 ns | 1.56 | 6.74 | 3.35 | 2.52 | 8.83 | 21.21 |
| 7 ns | 1.88 | 9.29 | 8.04 | 4.46 | 7.25 | 8.93 |
| 3 ns | 5.85 | 13 | 10.68 | 4.82 | 4.01 | 4.89 |
| 1 ns | 8.60 | 29.94 | 17.32 | 8.10 | 3.28 | 2.74 |
| 0.3 ns | 14.46 | 47.31 | 18.58 | 10.04 | 3.10 | 2.52 |
| 0.1 ns | 24.34 | 86.82 | 18.42 | 12.89 | 3.08 | 3.65 |

From Table 1, it is found that as the coincidence window decreases, both hyper-Poisson and super-Poisson states exhibit that the maximum FWHM monotonically decrease, while the standard deviation increases. Whereas, at larger coincidence windows, the peak amplitude becomes insufficient to clearly distinguish between the hyper-Poisson and super-Poisson states. As shown in Fig. 2 and Table 1, it can be said that a coincidence window of 1 ns provides a sufficiently sharp bunching peak in the autocorrelation function, enabling distinguish between these states while also significantly reducing noise. In the previous work [14] which used a 15 ns coincidence window, the autocorrelation function $g^2(0)$ exhibited a lower peak due to a weak bunching effect. As a result, the autocorrelation function $g^2(0)$ image was more susceptible to noise across the entire range from the minimum to the maximum values of autocorrelation function $g^2(0)$.

Demonstration of imaging

In previous work[14], the images were captured with a coincidence window of 15 ns. However, a major issue in the two-dimensional (2D) scanned image of the autocorrelation function was noise-induced fluctuation[14]. Additionally, as mentioned above, the resolution of image was insufficient. Accordingly, the spiral retarder sample (Fig 3(a)), similar to Ref. [14], was used to verify the effects of modification of coincidence window and introduction of objective lenses. Figure 3(b) shows the classical image obtained with a 1 ns coincidence window. The integration time of classical image (0.23 s) was adjusted so that the count level is similar to that of the coincidence image count level because the count rates are different between photon count rate ($R_A$) and coincidence rate. Figs. 3(c) and (d) show the 2D scan images of the coincidence and autocorrelation functions, respectively, with coincidence window (1 ns) and integration time (80 s). Note that the integration time longer than previous (20 s[14]) was given because the photon flux was significantly reduced due to the use of objective lenses.

In Fig. 3(c), the coincidence image shows enhanced contrast in regions corresponding to the slope of sinusoidal curve with respect to the azimuth angle, as reported in the reference[14]. Conversely, the regions at the maximum or minimum points of the sinusoidal curve, the contrast is reduced. In the autocorrelation function $g^2(0)$ image, the dark area, associated with the dip under the super-Poisson state, exhibit higher contrast. While the bright regions where the values of $g^2(0)$ are almost constant, which correspond to the region of the hyper-Poisson state, show weak and contrast. Due to the smaller change in the sample's polarization, the autocorrelation image provides better contrast than the coincidence or classical image.

Furthermore, the autocorrelation function effectively distinguishes the free space and the black plate, whereas the coincidence image fails to differentiate them. The autocorrelation function does not reach close to zero even at the minimum of the coincidence count rate or lower photon count rate of classical image. As a result, the autocorrelation function always starts from a non-zero baseline and includes an offset value, which in this case is less than 13%. Moreover using 1ns coincidence window, the autocorrelation function g² (0), exhibits a higher bunching peak. In the corresponding autocorrelation function g² (0) image, the noise level is confined only the narrower range up to the 10. Consequently, this results in a less noisy background image than that obtained with the 15 ns coincidence window[14]. Generally, at lower photon count area, the contrast in the coincidence image deteriorates, while the autocorrelation image maintains strong contrast, demonstrating its superior sensitivity under low-light conditions (in the order of Pico-joules).

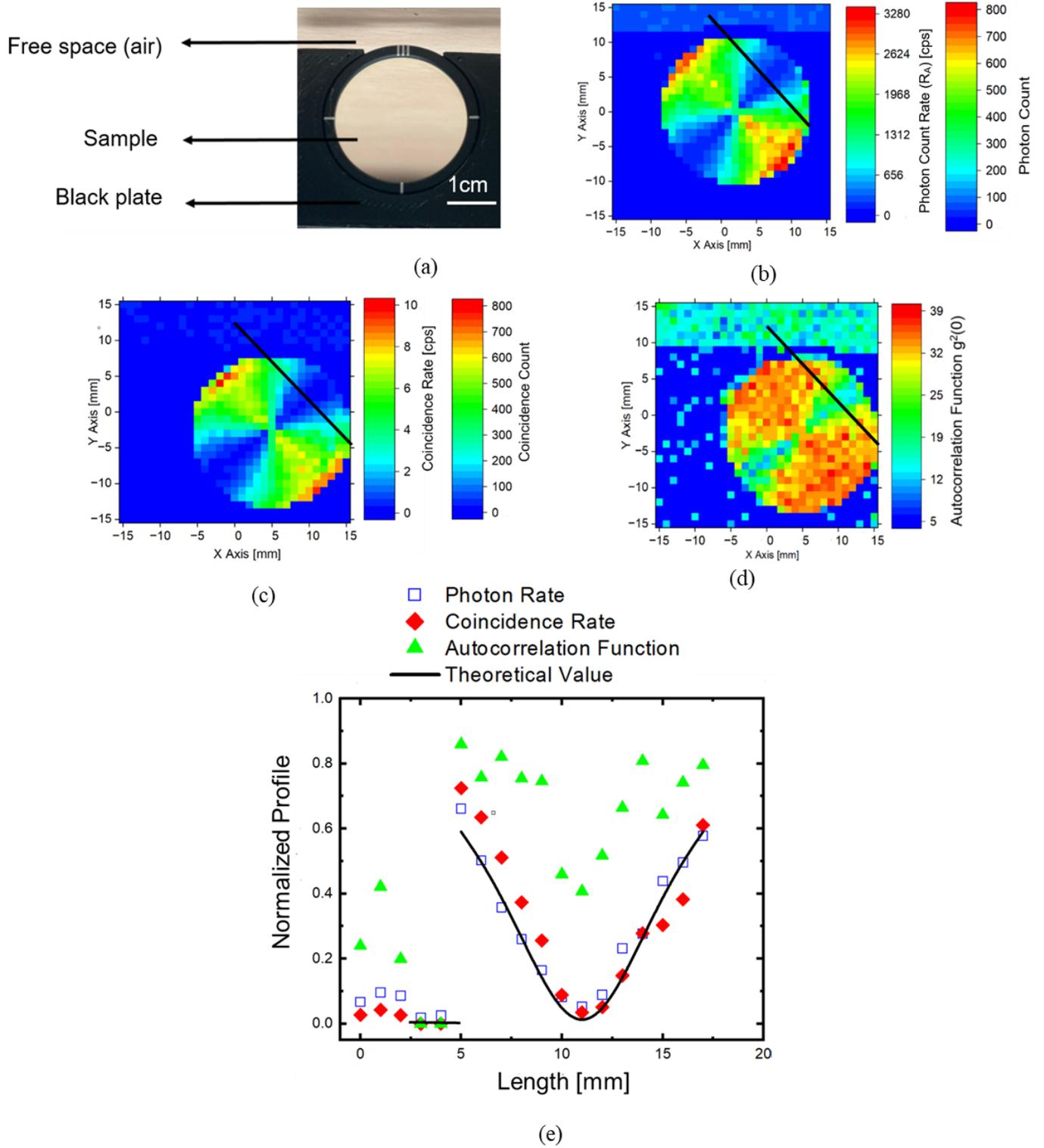

**Fig. 3.** Photograph of the optical spiral retarder sample[14)] (a); Classical image (b); Coincidence image (c); Autocorrelation image (d); and One-dimensional line-scanned profile highlighted in the black solid lines in the images (e).

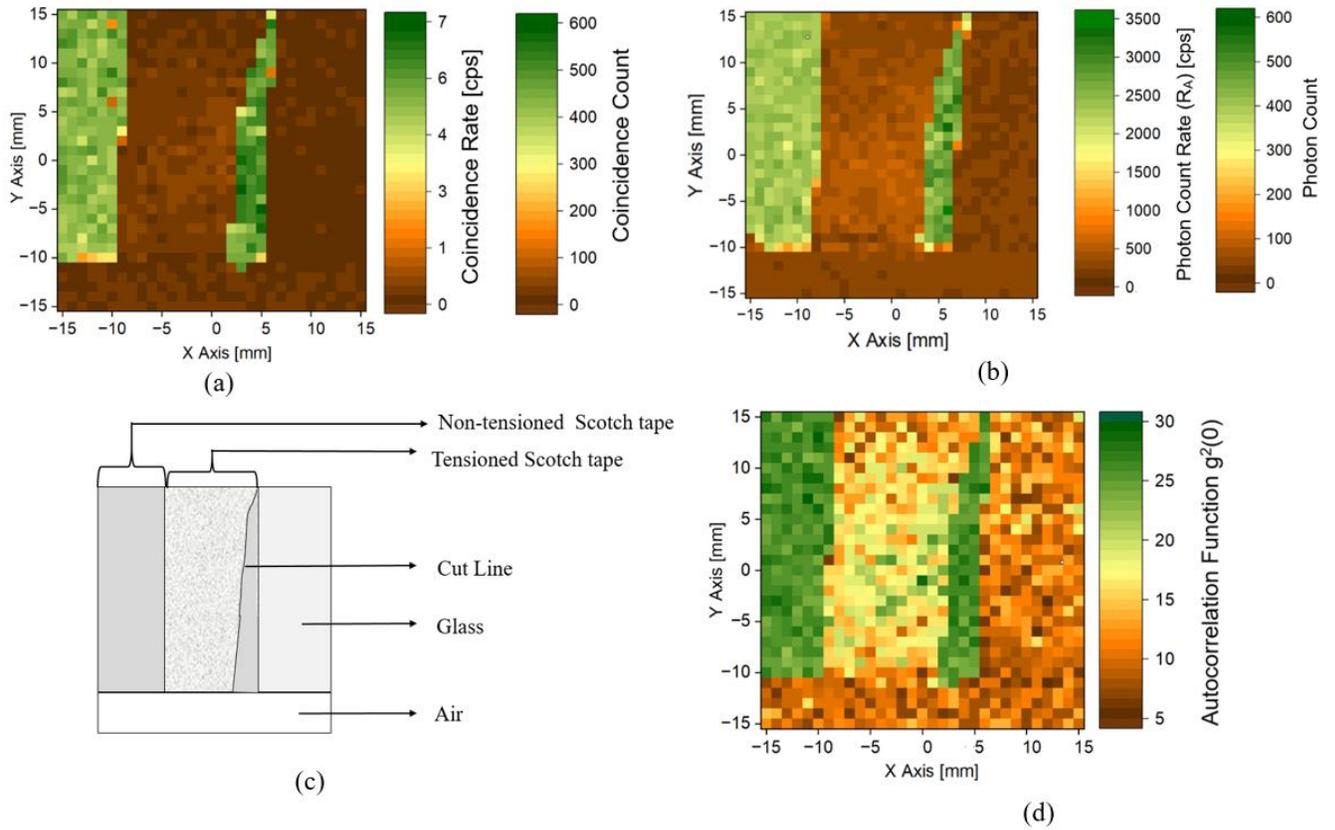

**Fig. 4.** Schematic view of scotch tape on glass plate sample (a); Classical image (b); Coincidence Image (c); and Autocorrelation image (d).

The black line in the figures, including different regions, such as free space, black plate, and sample areas, compares the observed images. Fig. 3 (e) exhibits the line profiles along the black solid line extracted from the photon rate, coincidence rate and autocorrelation function images.

As mentioned above, it was shown that the autocorrelation image shows strong contrast rather in the case of low photon count and small polarization variation. To investigate this more in detail, a sample consist of non-tensioned scotch tape, tensioned scotch tape, and a cut line on a glass plate (Fig. 4(a)) was prepared. Schematic view of the sample, the scanned classical, coincidence and autocorrelation images over a 30 × 30 mm area are shown in Fig. 4(a)–(d).

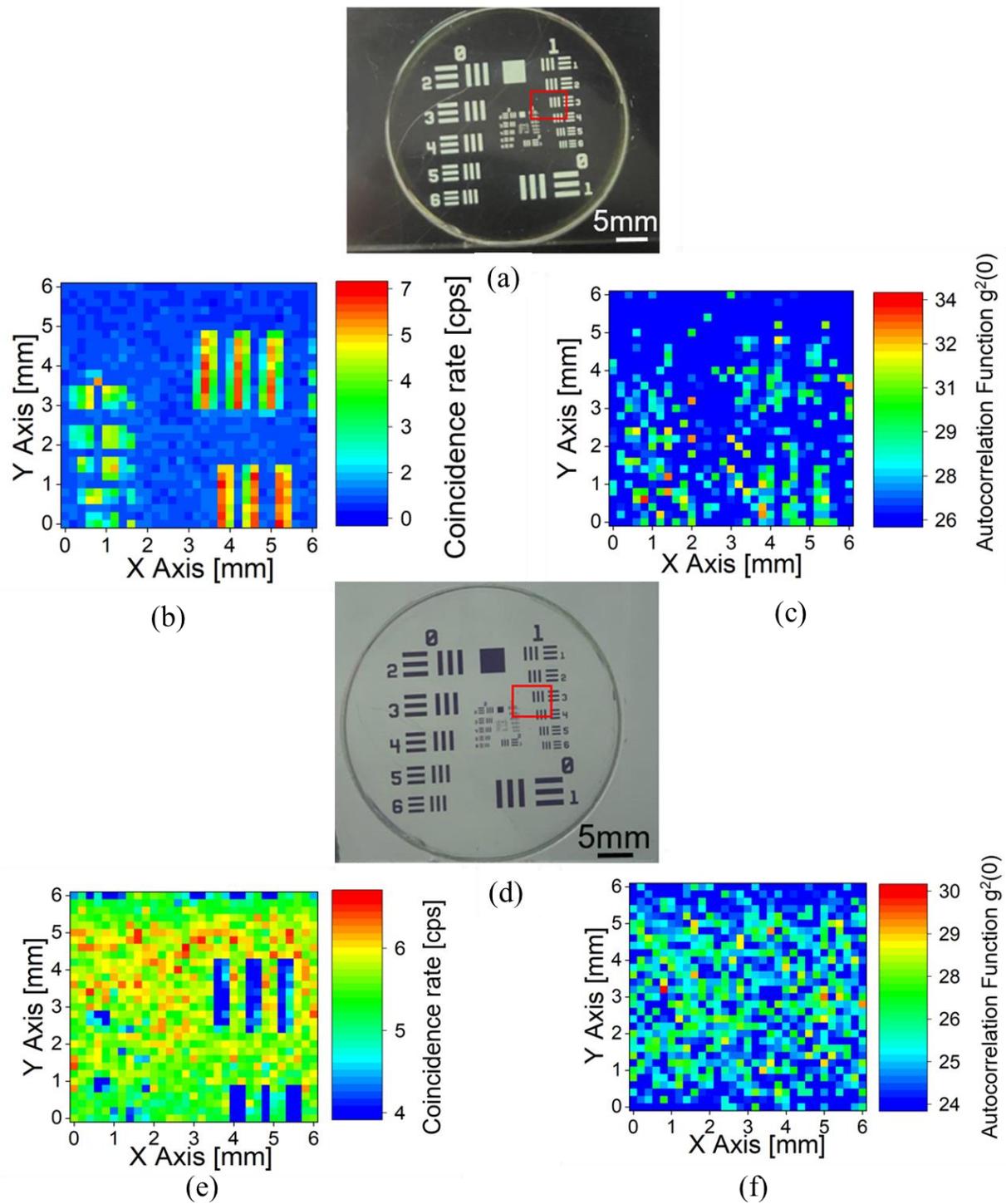

**Fig. 5.** Photograph of a USAF birefringence resolution target (Thorlabs R3L1S1B) sample as dark background and bright symbol (a); Coincidence image (b) and Autocorrelation function image (c) with the super-Poisson state at background; Photograph of the sample as bright background and dark symbol (d); Coincidence image (e) and Autocorrelation function image (f) with the hyper-Poisson state at background.

Again, note that the classical and coincidence images were obtained from the same count level adjusted by integration time though their count rates much different. The classical and coincidence images show that the regions corresponding to the tensioned Scotch tape, air medium, and glass plate cannot be distinguished. In contrast, the autocorrelation image clearly distinguishes the tensioned scotch tape region from the others. Therefore, in the case that change of polarization in the sample is small, the autocorrelation imaging is more favorable than ordinary polarization imaging.

A USAF birefringence resolution target (Thorlabs R3L1S1B) sample was employed to investigate the quality of our imaging method. Here, the sample was observed in the case that the background area (out of symbols) was made to dark condition and the symbol to bright region, as shown by the photograph in Fig. 5(a). In addition, the image of being inverted dark/bright region (Fig. 5(d)) was also observed. The resulting coincidence and autocorrelation images are shown in Fig. 5(b) and (c), and (e) and (f), respectively. In the coincidence image, the symbol is clearly visible, whereas in the autocorrelation image, the symbol is less distinct. Therefore, it was found that the autocorrelation image does not always show a fine contrast. In the case of this sample, photon scattering seemed to occur in the symbol region, resulting in a less contrast image, especially in the bright background configuration. Further details are provided in the supplementary material (see Fig. S1 and related text).

### 3.1 Discussion

Our imaging method is based on the transition between two different photon states (hyper-Poisson state and super-Poisson state) [14]. The statistics of photon emission are typically classified into three categories: coherent, Fock, and thermal states, each of which can be identified based on their photon number distribution[27]. In this sense, both the two different photon states used in our method are categorized into the thermal state. However, there should be some slight difference between these thermal states because the magnitude of bunching peak in the autocorrelation is clearly different. Meanwhile, one might infer that this peak height difference is due to that of photon count rate induced by blocking either the signal or idler port. In this section, we prove the transition between these two different photon states truly occurs by performing a specific photon statistical experiment.

To observe the transition between two different photon states by changing the analyzer angle, we performed Hanbury-Brown-Twiss (HBT) measurements with the SPDC entangled-photon source as a heralded single-photon source setup[28] (see Fig. S2 in the supplementary material). In this experiment, three detectors were employed, where the channel without beam splitter (directly connected from photon source to detector), and the channels with a beam splitter are denoted as Ch. T, and A and B, respectively. The coincidence count rates were recorded with an integration time of 100 s, in the case that no event, in which photons were coincidently detected at these three detectors, was aroused for 100 s, the measurements were maintained until 10 events occur.

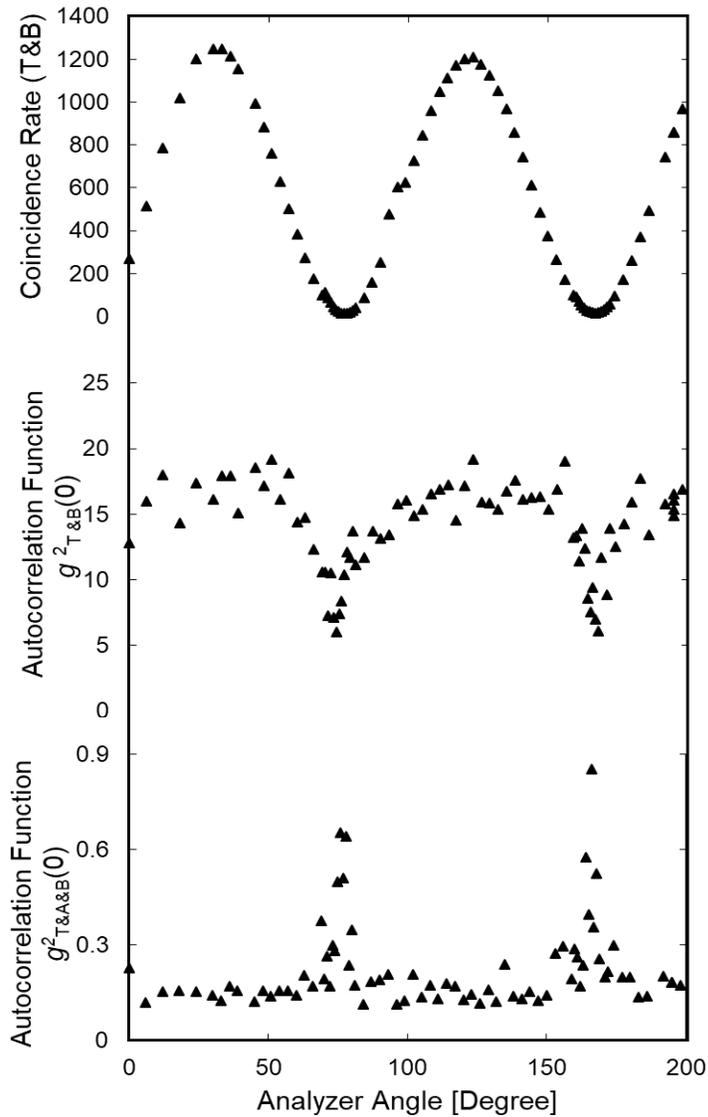

**Fig. 6.** Analyzer angle dependence of the coincidence rate (T&B), the second-order autocorrelation function of two channels [$g^2_{T\&B}(0)$], and the second-order autocorrelation function of three channels, [$g^2_{T\&A\&B}(0)$].

Figure 6 shows the analyzer angle dependence of the coincidence rate, the autocorrelation function between Ch. T and B, and the autocorrelation function among Ch. T, A and B. Naturally, the results from the two channel detectors were similar as those seen in our previous measurements[14]. In contrast, the three-channel autocorrelation $g^2_{T\&A\&B}(0)$ in the hyper-Poisson range consistently remained below 0.3, which is characteristic of a Fock state so-called photon antibunching[29]. On the other hand, the value of $g^2_{T\&A\&B}(0)$ in the super-Poisson range increased with decrease in the two-channel autocorrelation $g^2_{T\&B}(0)$ seen as a dip. Therefore, since the Fock and super-Poisson states are fundamentally different in nature[27, 29], this result supports the existence of distinct photon statistical states between the two distinct thermal states, i.e., hyper-Poisson and super-Poisson photon statistics.

Finally, we compare our method to the one reported in Ref.[30], which achieved biological imaging using spatial and polarization entanglement with an integration time of 1 s and a dynamic range of 34.84°. Although our method requires a longer integration time of 80 s, dynamic range of 29.56° is comparable.

Compared to the previous work using a 15 ns coincidence window, where the autocorrelation function image is much noisy, the background noise was fluctuating widely and reaching levels close to the maximum value of the autocorrelation function $g^2(0)$. Conversely, the current work using a 1 ns coincidence window demonstrates reduced noise and improved background clearness, because the noise was confined to a lower range. To sum up, the present work using the optimized coincidence window improved the image quality, i.e. a clearer background and less noise, better than the previous one[14].

## 4. Conclusion

In this study, we demonstrated an improved our quantum polarization microscopy technique utilizing an entangled-photon source. By optimizing a key experimental parameter, i.e. reducing the coincidence window to 1 ns, we effectively enhanced the transition between hyper-Poisson and super-Poisson states, thereby enabling accurate media identification with contrast even in the presence of noise. Additionally, incorporating an objective lens improved the resolution of image, addressing issues observed in the previous method. The transition between hyper-Poisson and super-Poisson states was surely evidenced by HBT measurements using a heralded single-photon source, revealing the distinct change in photon statistics. Furthermore, our study confirmed the advantages of using the autocorrelation function in imaging applications, particularly in distinguishing birefringent structures accurately. These improvements opened our proposed method as a promising approach for high-accuracy, extremely low-illumination imaging applications, particularly in biomedical and material science research.